\documentclass[appendixprefix=true]{article}
\usepackage[preprint]{spconf}
\usepackage{cite}
\usepackage{amsmath,amssymb,amsfonts}
\usepackage{array}
\usepackage{algorithmic}
\usepackage{graphicx}
\usepackage{textcomp}
\usepackage{xcolor}
\usepackage{bm}
\usepackage{multirow}
\usepackage{tikz}
\usepackage{mathtools}
\usepackage{hyperref}
\usepackage{cleveref}
\usepackage[ruled,vlined]{algorithm2e}
\usepackage{caption}

\copyrightnotice{\copyright\ IEEE 2023}
\toappear{Published in {\it Proc.\ ICASSP2023,
June 04-10, 2023, Rhodes Island, Greece}}

\newcommand{\IGNORE}[1]{}   

\newcommand{\ferdinand}[1]{\IGNORE{\textcolor[rgb]{0.13,0.54,0.54}
		{\textit{[Ferdinand: #1]}}}}

\newcommand{\floor}[1]{\left\lfloor #1\right \rfloor}
\def\TonalDeriv{0}  

\newcolumntype{x}[1]{>{\centering\arraybackslash\hspace{0pt}}p{#1}}

\title{Optimising Different Feature Types\\ 
 for Inpainting-based Image Representations}
\name{Ferdinand Jost \qquad Vassillen Chizhov \qquad Joachim Weickert
    \sthanks{This work has received funding from the European Research Council
	(ERC) under the European Union's Horizon 2020 research and
	innovation programme (grant agreement no. 741215, ERC Advanced
	Grant INCOVID).}}
\address{Mathematical Image Analysis Group, 
	Faculty of Mathematics and Computer Science\\
		Campus E1.7, Saarland University, 66041 Saarbr\"ucken, Germany\\
		\{jost, chizhov, weickert\}@mia.uni-saarland.de}

\usepackage{etoolbox}
\makeatletter
\patchcmd{\@@makechapterhead}{\endgraf\nobreak\vskip.5\baselineskip}{}{}{}
\makeatother

\begin{document}
\ninept
\maketitle

\begin{abstract}
Inpainting-based image compression is a promising alternative to classical
transform-based lossy codecs. Typically it stores a carefully selected 
subset of all pixel locations and their colour values. In the decoding phase
the missing information is reconstructed by an inpainting process such as
homogeneous diffusion inpainting. 
Optimising the stored data is the key for achieving good performance.
A few heuristic approaches also advocate alternative feature types such 
as derivative data and construct dedicated inpainting concepts. However, 
one still lacks a general approach that allows to optimise and inpaint 
the data simultaneously w.r.t.~a collection of different feature types, 
their locations, and their values. Our paper closes this gap.\\
We introduce a generalised inpainting process that can handle arbitrary 
features which can be expressed as linear equality constraints. This
includes e.g.~colour values and derivatives of any order. We propose a 
fully automatic algorithm that aims at finding the optimal features from 
a given collection as well as their locations and their function values 
within a specified total feature density. Its performance is demonstrated 
with a novel set of features that also includes local averages. Our 
experiments show that it clearly outperforms the popular inpainting with 
optimised colour data with the same density.
\end{abstract}

\section{Introduction}
\label{sec:introduction}

Inpainting is the process of reconstructing an image from a subset of 
its data \cite{GL14}. One of its most challenging applications is
lossy image compression. Inpainting-based codecs \cite{GWWB05} typically 
store a few well chosen pixel locations of the original image with 
their greyscale or colour values. In the decoding phase, the missing image 
parts are inpainted from these sparse data, often with a diffusion process. 
These methods have been able to outperform even widely used transform-based 
codecs such as JPEG and JPEG2000 \cite{SPME14}. Surprisingly, already the
simple linear process such as homogeneous diffusion inpainting can give 
good results, if the inpainting data is thoroughly optimised \cite{MHWT12}. 
This, however, is a highly nontrivial problem.
 
For achieving better visual quality, it has also been advocated to
replace the greyscale/colour data by gradient data \cite{BBG15,SPHW16}.  
However, these papers had to undertake various specific algorithmic 
adaptations, and the data optimisation problem becomes even harder.  
To further improve the quality, it has been suggested 
to combine the gradient information with greyscale/colour 
data \cite{SPHW16}. While this sounds promising, it has not been done 
so far. Moreover, it would be desirable to have a more general framework
that allows a straightforward incorporation of various classes of features
without the need for dedicated optimisation algorithms. 

\subsection{Our Contributions}

The goal of our paper is to address these challenges.
Our contributions are threefold:
\begin{enumerate}
\item We establish a generalised inpainting framework for linear
      inpainting operators that can handle any collection of features 
      in terms of linear equality constraints. This class is very large 
      and includes e.g.~derivatives of any order.
\item We introduce an efficient data selection strategy. It automatically 
      distributes the available data budget among all different features
      and optimises both their locations and their values. 
      This automates and generalises the otherwise cumbersome 
      feature-specific selection and optimisation process.
\item We identify a novel collection of features that includes local 
      averages. Experiments show that it considerably improves the 
      inpainting quality compared to classical inpainting.
\end{enumerate}
Since our paper focuses on feature integration and data optimisation,
we postpone any coding aspects to future work.

\subsection{Related Work}

Some inpainting-based codecs involve information at edges~\cite{Ca88,MBWF11} 
or isolines \cite{SCSA04}. However, these approaches still use grey values
as their only feature and just benefit from the fact that contours allow 
an inexpensive encoding of their locations.

Extensions of edge-like concepts that combine greyscale data with
the additional feature of discontinuities are presented in
\cite{HMWP13,JPW20,JPW21}. In contrast to our approach they use
specific segmentation concepts which do not generalise to other 
feature classes.

There are various attempts to reconstruct an image from features such 
as zero-crossings \cite{ZR86} or toppoints in scale-space \cite{KLDJ05},
as well as junctions \cite{CCM97}, and SIFT features \cite{WJP11}. While 
these papers give interesting information-theoretic insights, they do not 
offer competitive image representations in terms of compression quality.

Typically, an optimal placement of the features is crucial for the 
reconstruction quality. There has been 
a lot of work on spatial optimisation in the context of image 
inpainting, including analytic approaches \cite{BBBW08}, non-smooth 
optimisation \cite{HSW13, OCBP14, BLPP17}, neural networks
\cite{APW22}, probabilistic sparsification \cite{MHWT12}, and 
densification algorithms \cite{KBPW18,DAW21}. By combining the
ideas of error maps~\cite{KBPW18} and Voronoi densification~\cite{DAW21},
our approach falls into the latter class, but is the first one
to generalise it to large collections of feature types. 

\subsection{Paper Structure}

In \Cref{sec:our_framework} we review homogeneous diffusion inpainting,
we rewrite it as a constrained optimisation problem that covers many 
feature types, and we discuss efficient solution strategies for our novel, 
more general formulation. \Cref{sec:data_optimisation} introduces 
our strategy for optimising the feature locations and their values.
Finally, we evaluate the performance of our framework 
in~\Cref{sec:experiments}, and we give a conclusion and an outlook on 
future work in \Cref{sec:conclusion}.

\section{Our Framework}
\label{sec:our_framework}

In this section we give an overview of classical sparse inpainting with 
homogeneous diffusion and rewrite it in a variational formulation. 
This allows us to extend the problem and introduce any set of features 
that can be formulated as linear equations. We then suggest an efficient 
solution strategy for this generalised inpainting problem.

\subsection{Continuous Formulation}

Consider a continuous greyscale image 
$f(\bm{x}): \Omega \rightarrow \mathbb{R}$ where 
$\bm{x} := [x, y]^{\top}$ 
denotes a position in the rectangular image domain 
$\Omega \subset \mathbb{R}^2$. We assume that we have stored a sparse 
representation of $f$ only on the set of the \emph{inpainting mask} 
$K \subset \Omega$. A classical way \cite{Ca88} to inpaint the missing 
data is to compute a reconstruction $u : \Omega \rightarrow \mathbb{R}$ 
by solving the Laplace equation with Dirichlet conditions on the mask 
$K$ and reflecting boundary conditions on the domain boundary 
$\partial \Omega$:
\begin{equation}
    \label{eq:continuous_PDE_formulation}
    \begin{alignedat}{3}
    -\Delta u(\bm{x}) &= 0, &\quad &\bm{x}\in\Omega\setminus K,\\
                 u(\bm{x}) &= f(\bm{x}), &\quad &\bm{x} \in K, \\
    \partial_{\bm{n}}u(\bm{x}) &= 0, &\quad &\bm{x}\in\partial\Omega,
    \end{alignedat}
\end{equation}
where $\Delta=\partial_{xx}+\partial_{yy}$ is the Laplacian, and $\bm{n}$ 
denotes the normal vector to the boundary $\partial \Omega$. 
The fact that the Laplace equation $\Delta u = 0$ is the steady state 
of the homogeneous diffusion equation $\partial_t u=\Delta u$ \cite{Ii62}
motivates the name {\em homogeneous diffusion inpainting.} 
Problem (\ref{eq:continuous_PDE_formulation}) can be derived as the Euler-Lagrange 
equation of the variational formulation
\begin{equation}
\label{eq:continuous_variational_formulation}
\begin{gathered}
    \min_u\frac{1}{2}\int_{\Omega}\|\bm{\nabla} u(\bm{x})\|^2\,d\bm{x} 
    = \min_u \frac{1}{2}\int_{\Omega}u(\bm{x})(-\Delta) u(\bm{x})\,d\bm{x},\\ 
    \textrm{such that} \,\, u(\bm{x}) = f(\bm{x}), \,\, \bm{x}\in K,
\end{gathered}
\end{equation}
where $\|\cdot\|$ denotes the Euclidean norm, and
$\bm{\nabla}=[\partial_x,\partial_y]^\top$ is the nabla operator.
Here we have used the divergence theorem and the reflecting boundary
conditions. This formulation will provide a straightforward way to 
introduce other types of constraints.

\subsection{Discrete Formulation}

Digital images are typically represented on a regular pixel grid. Then the 
discrete analogue of $f$ can be represented as a vector 
$\bm{f}$ with dimension $N$ equal to the number of pixels. Similarly, 
we get the  
reconstruction vector $\bm{u}\in\mathbb{R}^N$. We also define the inpainting 
mask vector $\bm{c}\in\{0,1\}^N$ and its diagonal matrix 
$\bm{C}=\operatorname{diag}(\bm{c})$. This allows to discretise the
Dirichlet constraints $u(\bm{x})=f(\bm{x})$ on $K$ as 
$\bm{C}\bm{u} = \bm{C}\bm{f}$.  Finally we obtain the matrix 
$\bm{L} \in \mathbb{R}^{N \times N}$ from the standard 5-point stencil 
discretisation of the negated Laplacian ($-\Delta u_{i,j} \approx (-u_{i,j+1}-u_{i+1,j}+4u_{i,j}-u_{i-1,j}-u_{i,j-1})/h^2$) with reflecting boundary 
conditions. Putting everything together results in 
the discrete analogue to \Cref{eq:continuous_variational_formulation}:
\begin{equation}
\label{eq:discrete_inpainting_equation}
    \min_{\bm{u}}\frac{1}{2}\bm{u}^\top\bm{L}\bm{u}, \quad 
    \textrm{s.t.} \,\; \bm{C}\bm{u} = \bm{C}\bm{f}.
\end{equation}
Since $\bm{L}$ is a discretisation of $-\Delta$, it is a positive semidefinite 
matrix. Thus, the above is a special case of a quadratic programming 
problem with linear equality constraints~\cite{GMW81}. 
If the mask is 
non-empty, \Cref{eq:discrete_inpainting_equation} can be shown to have a 
unique solution that matches the unique solution of a direct discretisation 
of (\ref{eq:continuous_PDE_formulation}); see also \cite{MBWF11}. 

\subsection{Generalised Discrete Formulation}

Our goal is to extend the discrete inpainting in 
\Cref{eq:discrete_inpainting_equation} to not only consider grey values as 
constraints, but a collection of features that can be implemented through 
linear equality constraints. To this end, we replace the Dirichlet 
constraints $\bm{Cu} = \bm{Cf}$ by $m$ types of constraints of the form 
$\bm{C}_i\bm{A}_i\bm{u} = \bm{C}_i\bm{A}_i\bm{f}$ with 
$i \in \{1, \ldots, m\}$:
\begin{equation}
    \label{eq:formulation_compat_constraints}
    \begin{gathered}
    \min_{\bm{u}}\frac{1}{2}\bm{u}^\top\bm{L}\bm{u}, \quad 
    	\textrm{s.t.} \,\, \bm{A}\bm{u} = \bm{b}\\[1mm]
    \bm{A} := \begin{bmatrix}\bm{C}_1\bm{A}_1\\ \ldots \\ 
    	\bm{C}_m\bm{A}_m\end{bmatrix}, \quad
    \bm{b} := \begin{bmatrix} \bm{C}_1\bm{A}_1\bm{f} \\ \ldots \\ 
    	\bm{C}_m\bm{A}_m\bm{f}\end{bmatrix}.
    \end{gathered}
\end{equation}
The matrices $\bm{A}_i \in \mathbb{R}^{N \times N}$ describe convolutions 
with user-defined feature stencils. For example, we can implement Dirichlet 
constraints through $\bm{A}_1 = \bm{I}$ with identity matrix $\bm{I}$.
First-order derivative constraints can be modelled by $\bm{A}_2=\bm{D}_x$ 
and $\bm{A}_3=\bm{D}_y$ with forward difference matrices  $\bm{D}_x,\, 
\bm{D}_y$ that approximate $\partial_x, \, \partial_y$. Also local
integral constraints can be included easily, since they are linear
operators as well. 
We also note that the above formulation is rather general w.r.t.~the 
discrete linear inpainting operator $\bm{L}$: It can be any symmetric 
positive semidefinite matrix, e.g.~a discretisation of the 
biharmonic operator $\Delta^2$. 
Considering linear inpainting operators and linear constraints keeps 
our discussion simple and is not very limiting, since they can give good 
reconstructions, if the data is optimised carefully \cite{MHWT12}.
Extending this inpainting to colour images is straightforward: We can 
treat each channel separately.

\subsection{Numerical Solution Strategy}
\label{sec:optimisation}

The constrained optimisation problem (\ref{eq:formulation_compat_constraints}) 
can be solved with a Lagrange multiplier approach \cite{GMW81}. This 
turns it into an unconstrained one by introducing an additional vector 
$\bm{\lambda} \in \mathbb{R}^{m N}$ of unknowns: 
%
\begin{equation}
	\begin{gathered}
	\min_{\bm{u}}\max_{\bm{\lambda}}\left[\frac{1}{2}\bm{u}^{\top}\bm{L}\bm{u} 
		+ \bm{\lambda}^{\top}(\bm{A}\bm{u}-\bm{b})\right].
    \end{gathered}
\end{equation}
%
Setting the derivatives w.r.t.~$\bm{u}$ and $\bm{\lambda}$ to zero gives
the linear system
\begin{equation}
\begin{gathered}
    \label{eq:Lagrange_equation}
    \begin{bmatrix} 
    	\bm{L} & \bm{A}^{\top} \\ \bm{A} & \bm{0} \end{bmatrix} 
    	\begin{bmatrix} \bm{u} \\ \bm{\lambda} \end{bmatrix} 
    	= \begin{bmatrix} \bm{0} \\ \bm{b} \end{bmatrix}.
    \end{gathered}
\end{equation}
Its system matrix is symmetric but indefinite. After evaluating a variety 
of solvers, our algorithm of choice is SYMMLQ \cite{PS75}. Similar to the 
conjugate gradients solver~\cite{Sa03} it can be derived from the Lanczos 
method. However, unlike conjugate gradients, it is able to handle indefinite 
matrices. In fact, it can even handle singular systems as long as those 
have a solution. This is useful in our setting, since it even allows for 
features that are linearly dependent.

\section{Data Optimisation}
\label{sec:data_optimisation}

In order to achieve a good reconstruction quality with small error 
$\|\bm{u}-\bm{f}\|^2$, one must optimise the feature masks 
$\bm{C}_i$ and the stored feature values $\bm{b}$ in 
\Cref{eq:formulation_compat_constraints}. Thus, let us now propose an 
efficient and generic 
\emph{spatial optimisation} algorithm for the masks $\bm{C}_i$, as well as 
a \emph{tonal optimisation} method for the feature values $\bm{b}$.


\subsection{Spatial Optimisation}
\label{sec:spatial_optimisation}
Our spatial optimisation algorithm combines error maps~\cite{KBPW18} and 
a Voronoi densification~\cite{DAW21} approach to efficiently construct 
the inpainting masks $\bm{C}_1,\ldots,\bm{C}_m$ for the features 
$\bm{A}_1,\ldots,\bm{A}_m$. Given a target number of mask points $|K|$,  
it incrementally constructs the masks in $n$ iterations (requiring $n$ 
inpaintings). In each iteration, it introduces $\floor{|K|/n}$ mask points. More 
iterations require a longer runtime, but offer a better quality. 

We want to find good locations for the $\floor{|K|/n}$ mask points to 
be inserted, using a minimal number of inpaintings. To this end, we 
use the inpainting $\bm{u}$ from the previous iteration, and we 
compute $m$ error maps 
$|\bm{A}_i(\bm{u}-\bm{f})|^2 \in\mathbb{R}^N, \, i\in \{1,\ldots,m\}$. 
The absolute value squared $|\cdot|^2$ is taken pointwise (in the colour case 
this is the Euclidean norm over RGB vectors); see Fig.~\ref{fig:exp_densification}, 
column 2 for an illustration. This expression allows us to 
minimise the mean squared error (MSE). The multiplication of the signed 
error $\bm{u}\!-\!\bm{f}$ with $\bm{A}_i$ serves to capture the feature-wise 
error. 

A straightforward application of the above error maps would be to insert mask 
points at the locations of highest pointwise error. However, this does not 
yield the best possible results, as it does not reflect the reduction of the 
error in the neighbourhood of the inserted points. As a simple approximation
to capture this effect locally, we partition the image with the Voronoi 
decomposition induced by the current set of mask points. Then we integrate 
the $m$ error maps in each cell, which gives $m$ error values. 
Each cell is assigned its highest integrated error and the corresponding 
feature type. We then insert 
one mask point per cell in the $\floor{|K|/n}$ cells with largest error. The points 
within each cell are of the feature type of that cell. They are inserted 
at the location of the highest pointwise error in that cell.
For an illustration, see \Cref{fig:exp_densification}.

\subsection{Tonal Optimisation}
\label{sec:tonal_optimisation}
After the spatial optimisation, we can also optimise the values $\bm{b}$ 
at the mask points. Since (\ref{eq:Lagrange_equation}) is a system of 
linear equations, the inpainting part $\bm{u}$ of its solution can be 
written as $\bm{u}=\bm{R}\,\bm{b}$ with some reconstruction matrix $\bm{R}$. 
Our tonal optimisation problem then reads
\begin{align}
   \min_{\bm{b}}\|\bm{u}-\bm{f}\|^2 
   &= \min_{\bm{b}}\|\bm{R}\,\bm{b}-\bm{f}\|^2.
\end{align}
It generalises the classical tonal optimisation of grey values \cite{MHWT12} 
to a tonal optimisation of various feature types. 
If we set the derivative w.r.t.~$\bm{b}$ to zero, 
we obtain the normal equations
\begin{align}
    \bm{R}^{\top}\!\bm{R}\,\bm{b}  &= \bm{R}^{\top}\bm{f}.
\end{align}
We solve these normal equations with the CGNR algorithm~\cite{Sa03}. 
For efficiency purposes we do not explicitly compute $\bm{R}$, since it 
is a dense matrix. Instead we use SYMMLQ \cite{PS75} to evaluate the 
matrix--vector products with $\bm{R}$ and $\bm{R}^{\top}$ in 
the CGNR algorithm.
\if\TonalDeriv1
The specific form of $\bm{R}$ can be derived from \Cref{eq:Lagrange_equation} 
as follows:
\begin{align}
\begin{split}
    \begin{bmatrix} \bm{L} & \bm{A}^T \\ \bm{A} & \bm{0} \end{bmatrix} 
    \begin{bmatrix} \bm{u} \\ \bm{\lambda} \end{bmatrix} 
    &= \begin{bmatrix} \bm{0} \\ \bm{b} \end{bmatrix} 
    \\
    \begin{bmatrix} \bm{u} \\ \bm{\lambda} \end{bmatrix} 
    &=  \begin{bmatrix} \bm{L} & \bm{A}^T \\ \bm{A} & \bm{0} \end{bmatrix}^{-1}
    \begin{bmatrix}\bm{0} \\ \bm{I}\end{bmatrix} \bm{b} 
    \\
    \bm{u} =\bm{R}\bm{b} &= \begin{bmatrix} \bm{I} & \bm{0}\end{bmatrix}
    \begin{bmatrix} \bm{L} & \bm{A}^T \\ \bm{A} & \bm{0} \end{bmatrix}^{-1}
    \begin{bmatrix}\bm{0} \\ \bm{I}\end{bmatrix}\bm{b}
\end{split}
\end{align}
Then the multiplication by $\bm{R}$ and $\bm{R}^\top$ in the CGNR iterations 
can be written as a prolongation, an inpainting (the multiplication by the 
inverse matrix), and a restriction.
\fi

\section{Experiments}

\label{sec:experiments}

We illustrate the qualitative benefit of our framework on two natural 
colour images of size $512 \times 512$: {\em elpaso} and {\em windmill}
(photos by J.~Weickert). 
Our proposed example features consist of colour values, forward 
differences in $x$- and $y$-direction, and local colour averages on 
$2 \times 2$ and $16 \times 16$ patches, respectively. 

\Cref{fig:exp_qualitative} studies the influence of the number of feature
types in a sparse representation with a combined mask density of $5\%$.
The masks are optimised with 30 iterations of our proposed Voronoi 
densification. Increasing the variety of features -- while keeping the
same total mask density -- decreases the reconstruction error in a 
monotone way. This shows that all proposed feature types are beneficial,
and that our optimisation strategy over the different feature types
is effective in practice. It should be noted that using five feature 
types instead of a single one (the classical colour values) comes at 
basically no extra expense.

\Cref{fig:exp_tonal_optimisation} illustrates the impact of our
generalised tonal optimisation. We observe that applying it to the 
spatially optimised representations with five feature types improves 
the MSE by about one third. Also here one should note that optimising 
the function values does not increase the total amount of data: Good 
values cost as much as bad ones.
 

\setcounter{figure}{1}
\begin{figure}
    \centering
    \begin{tabular}{ccc}
        original & without tonal opt. & with tonal opt.\\ 
        \includegraphics[width=0.135\textwidth]
        	{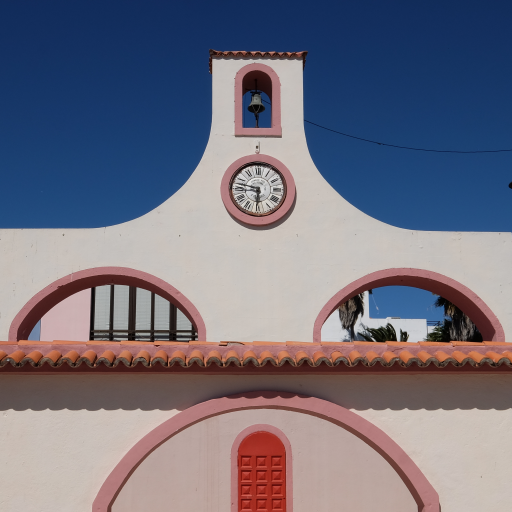} & 
        \includegraphics[width=0.135\textwidth]
            {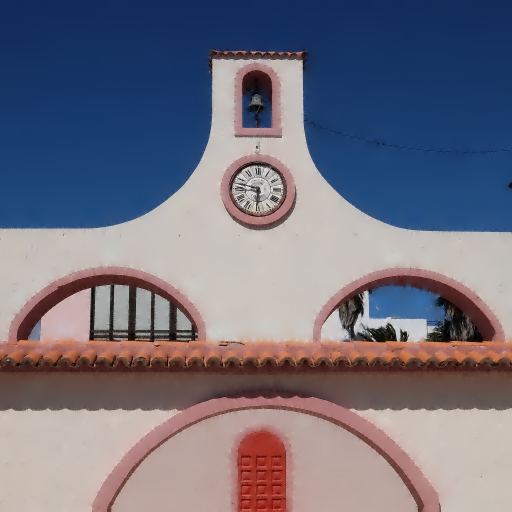} &
        \includegraphics[width=0.135\textwidth]
        	{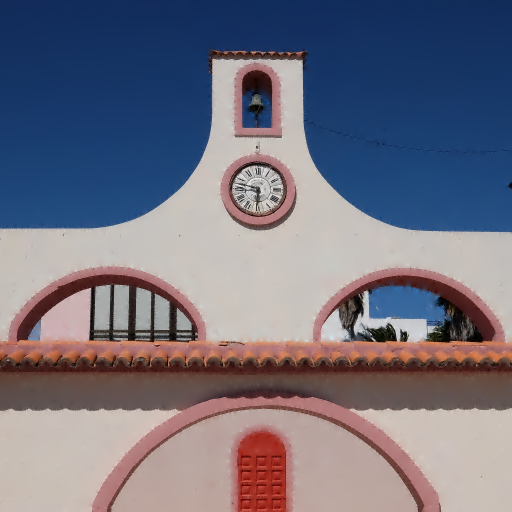}\\ 
        {\em elpaso} & MSE: $23.25$ & MSE: $\bm{15.79}$ \\[2mm]
        
        \includegraphics[width=0.135\textwidth]
        	{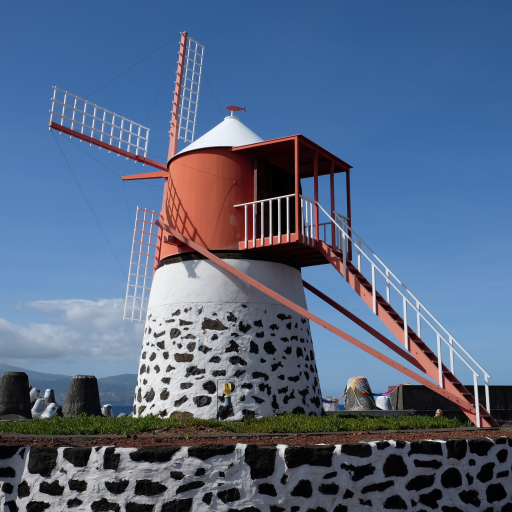} & 
        \includegraphics[width=0.135\textwidth]
        	{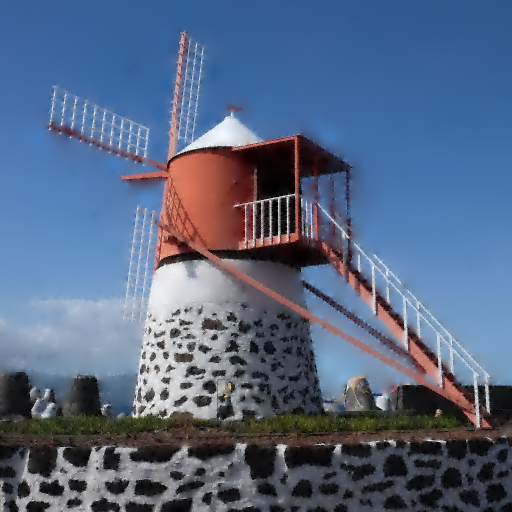} &
        \includegraphics[width=0.135\textwidth]
        	{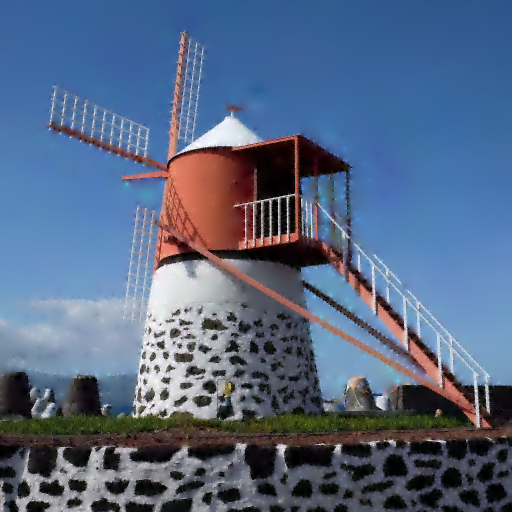} \\ 
        {\em windmill} & MSE: $157.30$ & MSE: $\bm{108.54}$\\
    \end{tabular}
    \caption{Test images \emph{elpaso} and \emph{windmill} and their
        sparse representations using all five proposed feature types 
        (total mask density: 5 \%) without and with tonal optimisation. 
        The tonal optimisation improves the reconstruction quality by 
        about one third.}
    \label{fig:exp_tonal_optimisation}
\end{figure}

\setcounter{figure}{0}
\begin{figure*}
\center
    \begin{tabular}{c c c c}
    Input & Errors & Integrated Errors & Output \\
    \includegraphics[width=0.14\textwidth]
    	{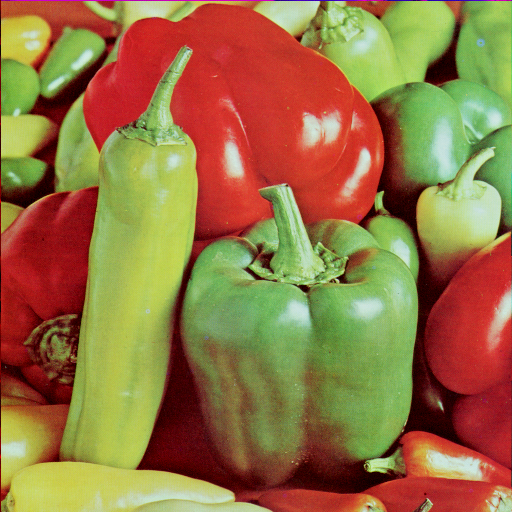} & 
    \includegraphics[width=0.14\textwidth]
    	{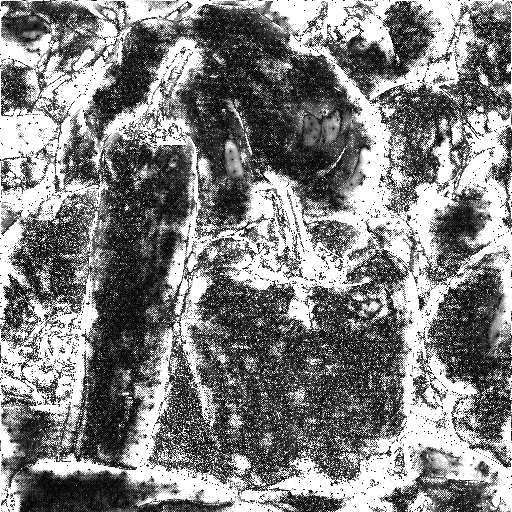} &
    \includegraphics[width=0.14\textwidth]
    	{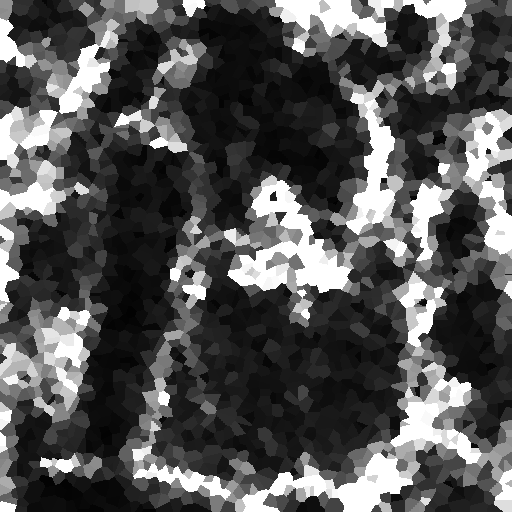} &
    \includegraphics[width=0.14\textwidth]
    	{resources/dens/peppers.png}
    \\
    \includegraphics[width=0.14\textwidth]
    	{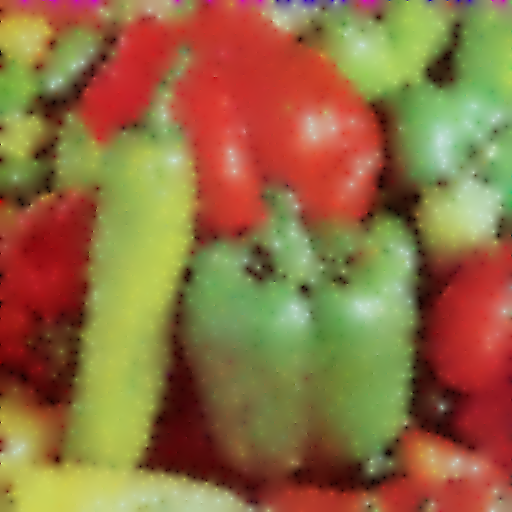} & 
    \includegraphics[width=0.14\textwidth]
    	{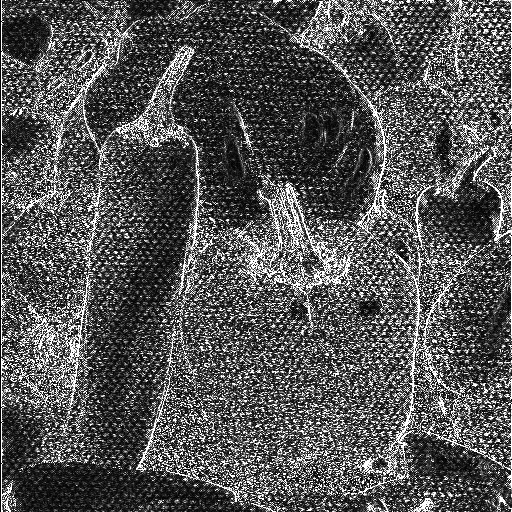} &
    \includegraphics[width=0.14\textwidth]
    	{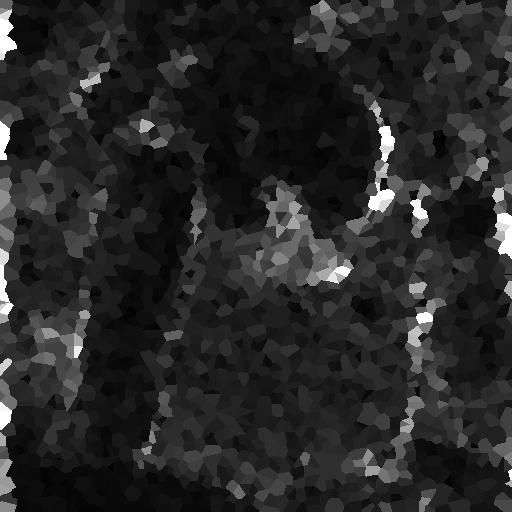} &
    \includegraphics[width=0.14\textwidth]
    	{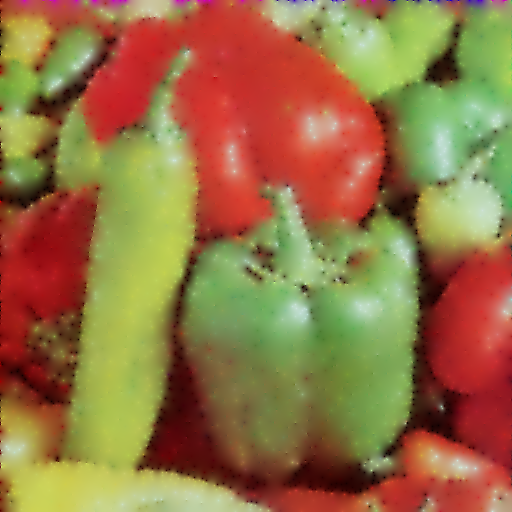}
    \\
    \includegraphics[width=0.14\textwidth]
    	{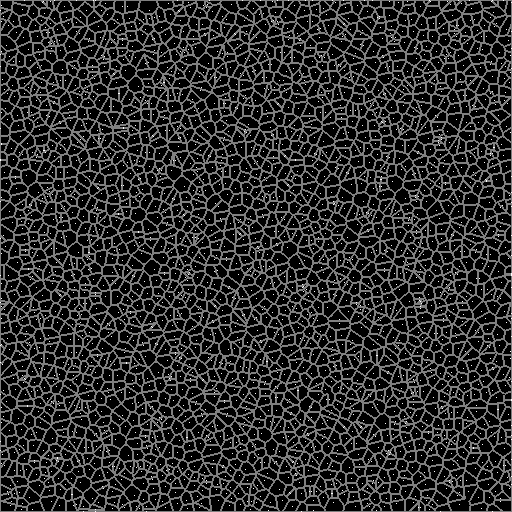} & 
    \includegraphics[width=0.14\textwidth]
    	{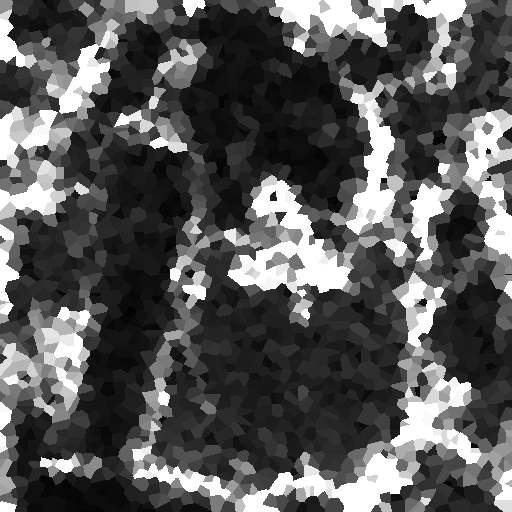} &
    \includegraphics[width=0.14\textwidth]
    	{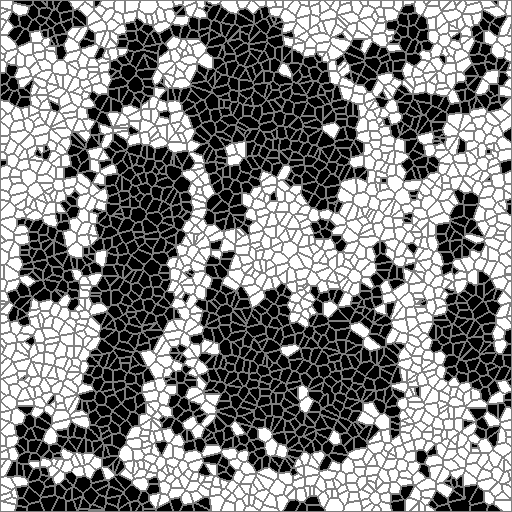} &
    \includegraphics[width=0.14\textwidth]
    	{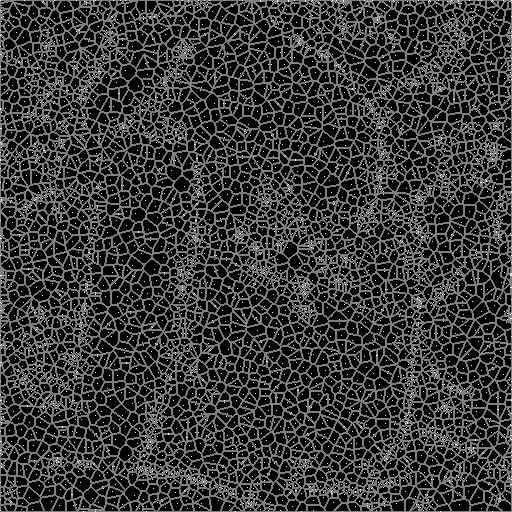}
    \end{tabular}
    \caption{Steps in a single iteration of our densification algorithm. 
    	The {\bf first column} is the output of the previous step. From top
        to bottom: reference image, inpainting, mask with Voronoi 
        diagram. The {\bf second column} consists of 
    	the error maps for the Dirichlet and $\partial_x$ features, and the 
    	maximum cell-wise integrated errors. The {\bf third column} contains 
    	cell-wise integrated errors for the Dirichlet and $\partial_x$ 
    	features, and the selected $\floor{|K|/n}$ cells with largest errors that 
    	are to be refined (in white). The {\bf last column} shows the output 
        of this iteration: reference image, new inpainting, and updated mask 
        with Voronoi diagram.}
    \label{fig:exp_densification}
\end{figure*}

\setcounter{figure}{2}
\begin{figure*}
\center
    {
    \setlength{\tabcolsep}{1pt}
    \begin{tabular}{c x{0.11\textwidth} x{0.11\textwidth} x{0.11\textwidth} 
        x{0.11\textwidth}@{\hskip 10pt} x{0.11\textwidth} x{0.11\textwidth} 
        x{0.11\textwidth} x{0.11\textwidth}}

    	\multirow{1}{*}[8ex]{\rotatebox[origin=c]{90}{inpainting}} &
    	\includegraphics[width=0.09\textwidth]
    		{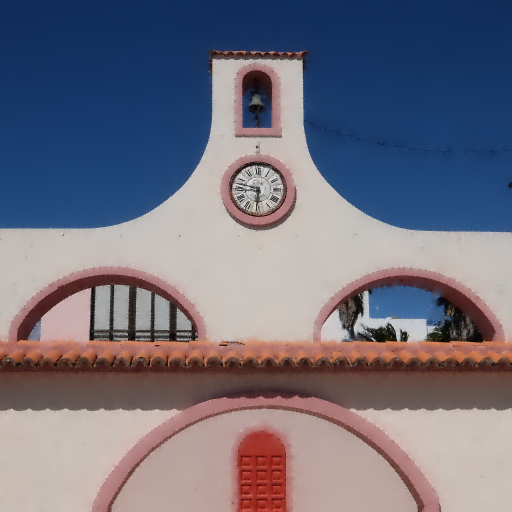} &
    	\includegraphics[width=0.09\textwidth]
    		{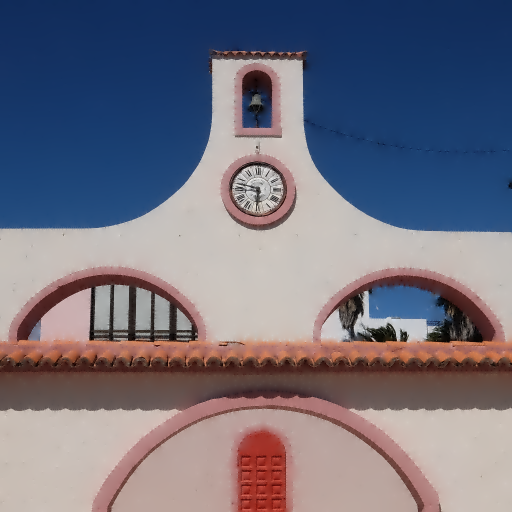} &
    	\includegraphics[width=0.09\textwidth]
    		{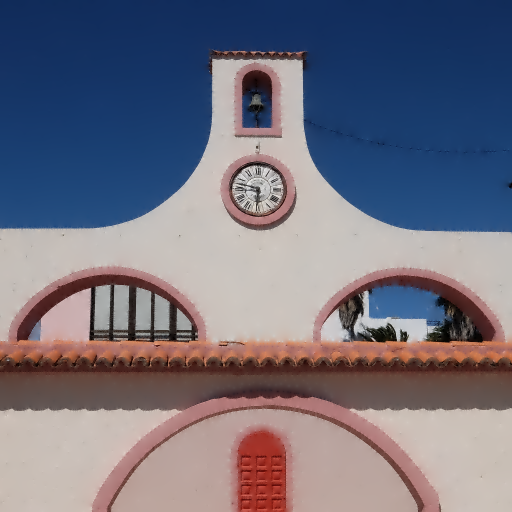} &
    	\includegraphics[width=0.09\textwidth]
    		{resources/elpaso/elpaso-icassp_5_050000.png} &
    	
    	\includegraphics[width=0.09\textwidth]
    		{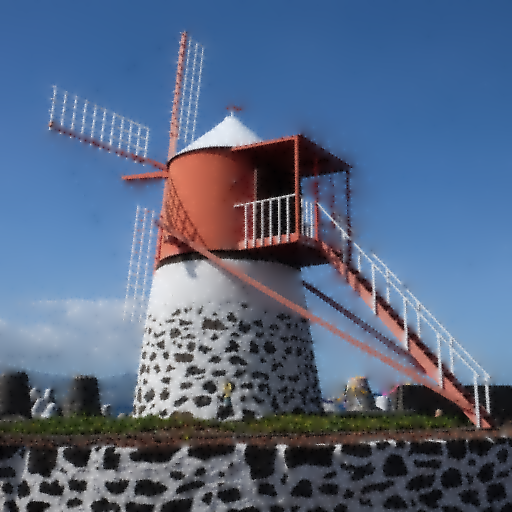} &
    	\includegraphics[width=0.09\textwidth]
    		{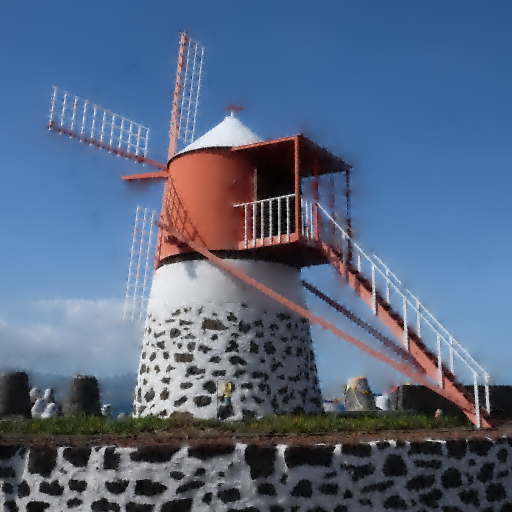} &
    	\includegraphics[width=0.09\textwidth]
    		{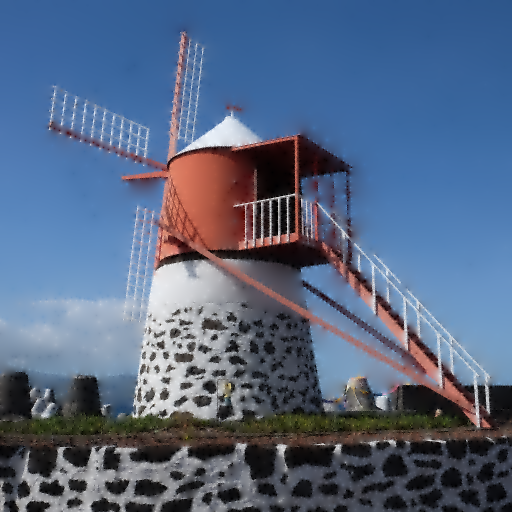} &
    	\includegraphics[width=0.09\textwidth]
    		{resources/windmill/windmill-icassp_5_050000.png} \\
    	
    	\multirow{1}{*}[6ex]{\rotatebox[origin=c]{90}{colour}} &
    	\includegraphics[width=0.09\textwidth]
    		{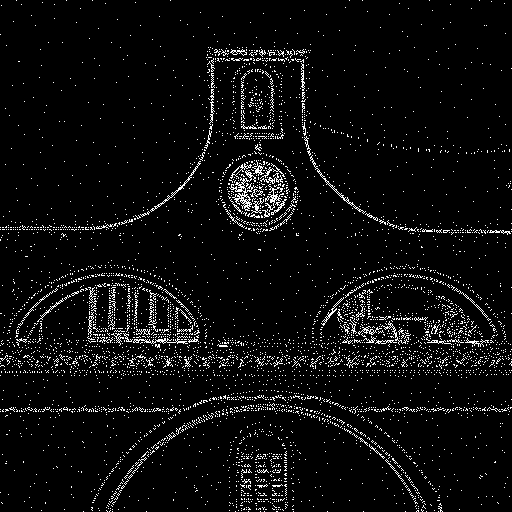} &
    	\includegraphics[width=0.09\textwidth]
    		{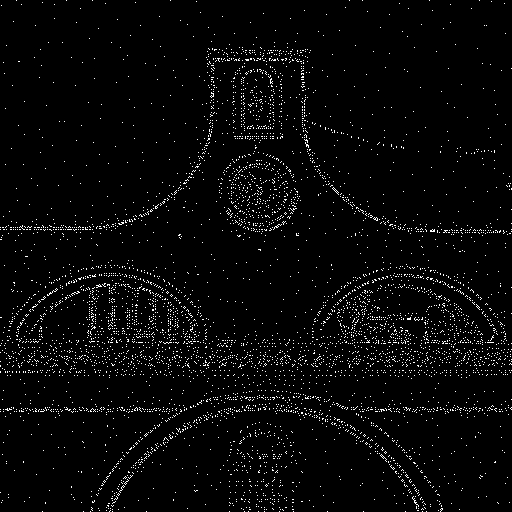} &
    	\includegraphics[width=0.09\textwidth]
    		{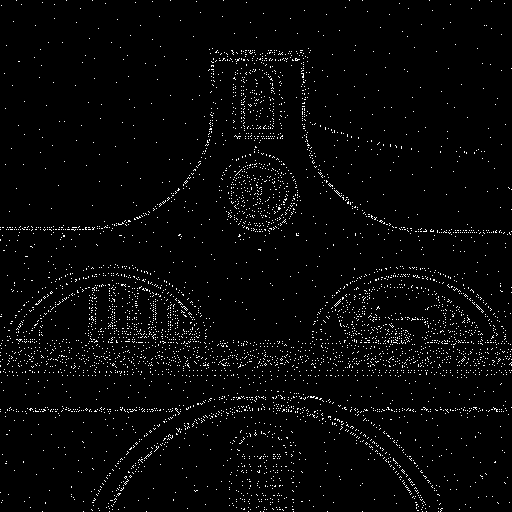} &
    	\includegraphics[width=0.09\textwidth]
    		{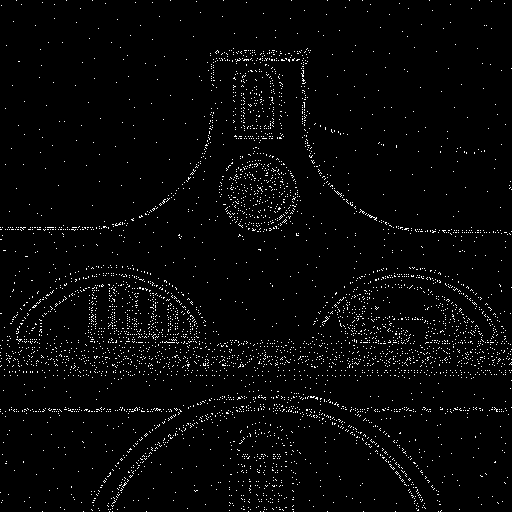} &
    	
    	\includegraphics[width=0.09\textwidth]
    		{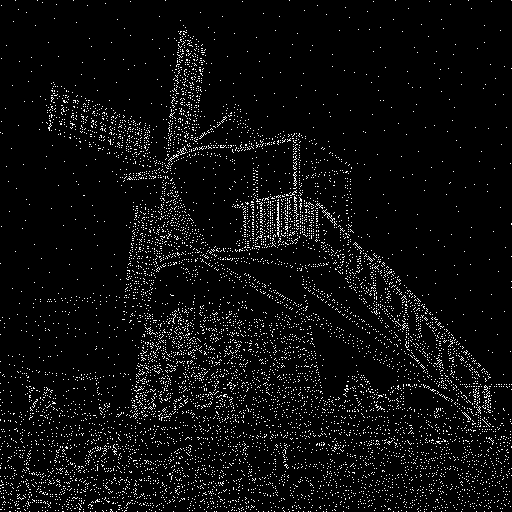} &
    	\includegraphics[width=0.09\textwidth]
    		{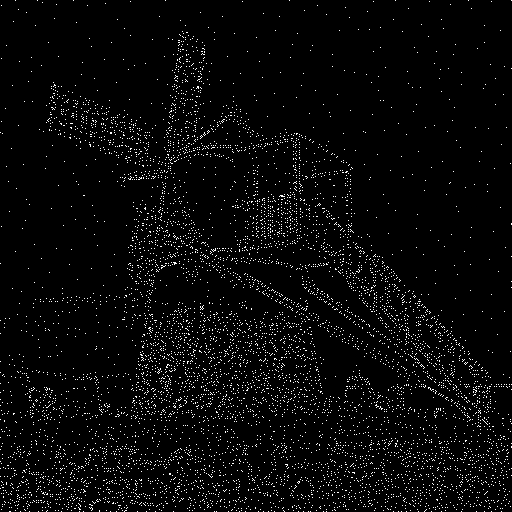} &
    	\includegraphics[width=0.09\textwidth]
    		{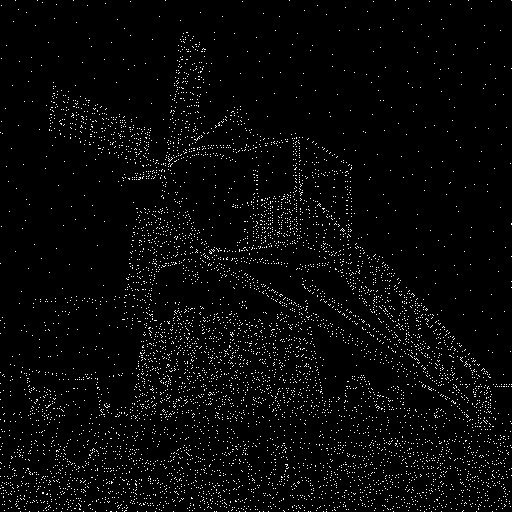} &
    	\includegraphics[width=0.09\textwidth]
    		{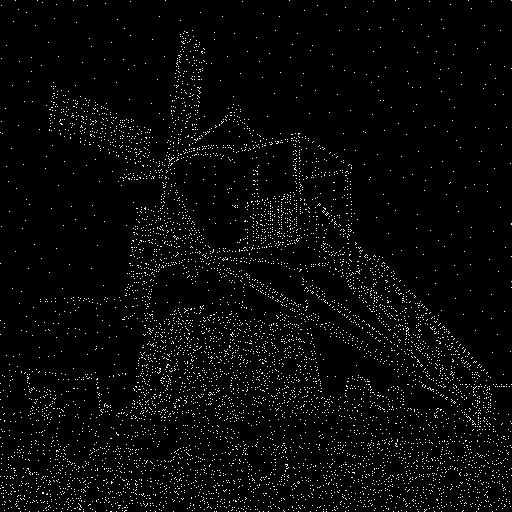} \\
    	
    	\multirow{1}{*}[5ex]{\rotatebox[origin=c]{90}{$\partial_x$}} &
    	~ &
    	\includegraphics[width=0.09\textwidth]
    		{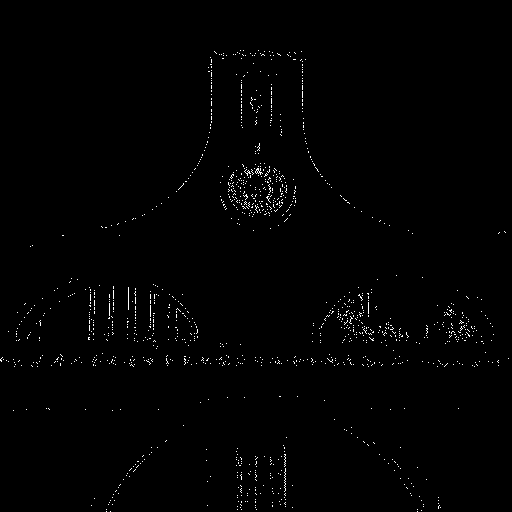} &
    	\includegraphics[width=0.09\textwidth]
    		{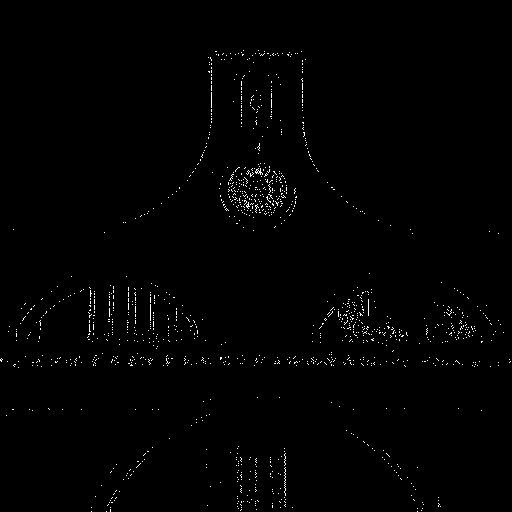} &
    	\includegraphics[width=0.09\textwidth]
    		{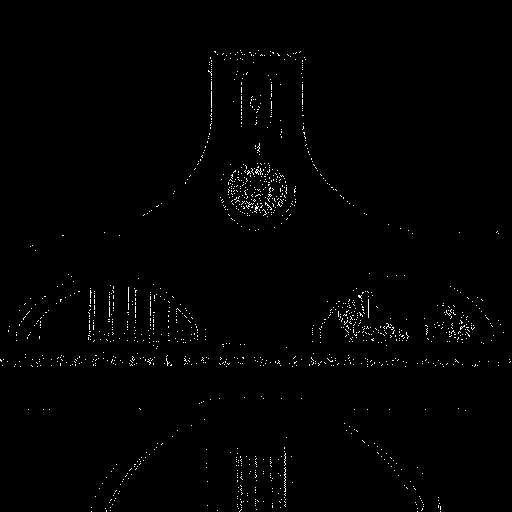} &
    	
    	~ &
    	\includegraphics[width=0.09\textwidth]
    		{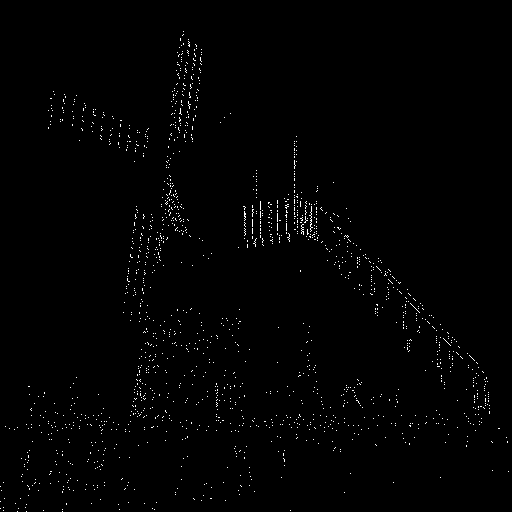} &
    	\includegraphics[width=0.09\textwidth]
    		{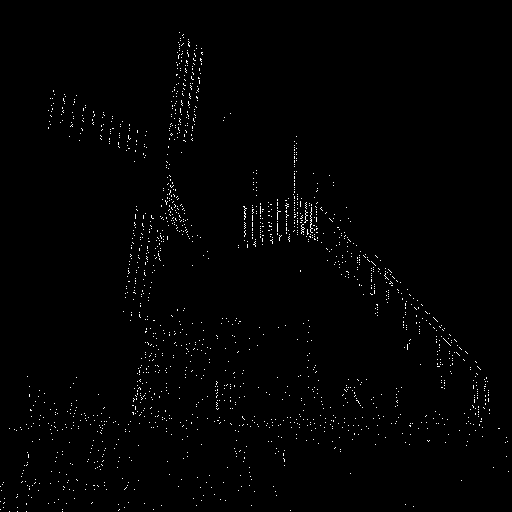} &
    	\includegraphics[width=0.09\textwidth]
    		{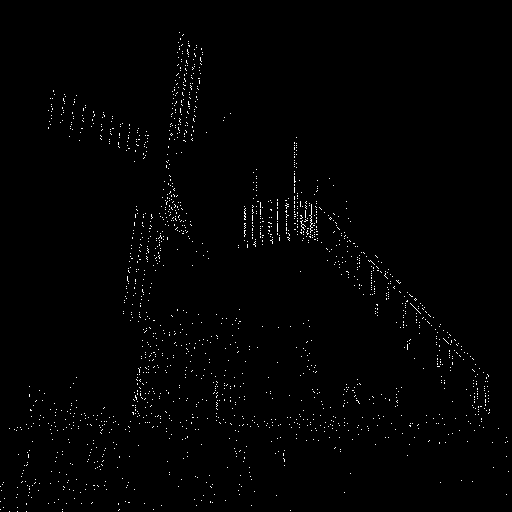} \\
    	
    	\multirow{1}{*}[5ex]{\rotatebox[origin=c]{90}{$\partial_y$}} &
    	~ &
    	\includegraphics[width=0.09\textwidth]
    		{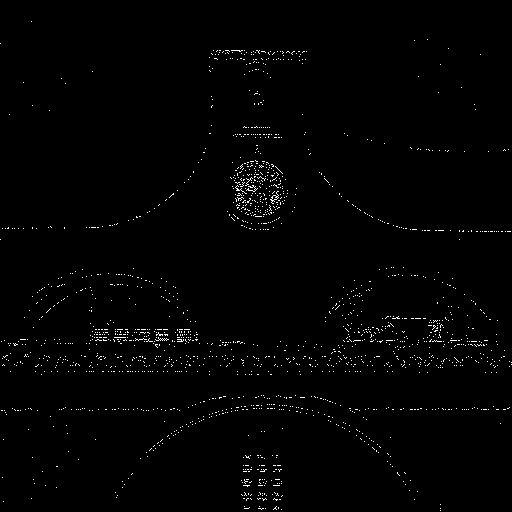} &
    	\includegraphics[width=0.09\textwidth]
    		{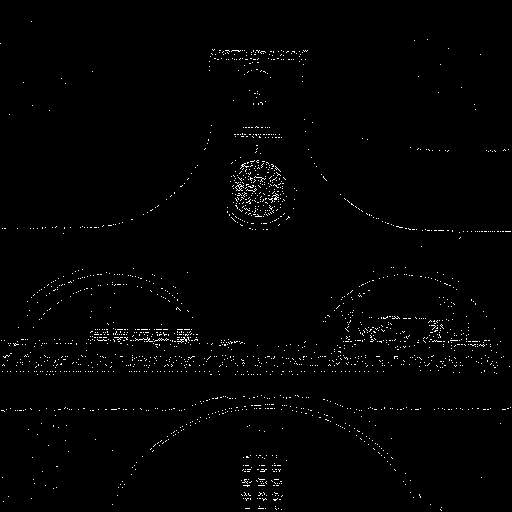} &
    	\includegraphics[width=0.09\textwidth]
    		{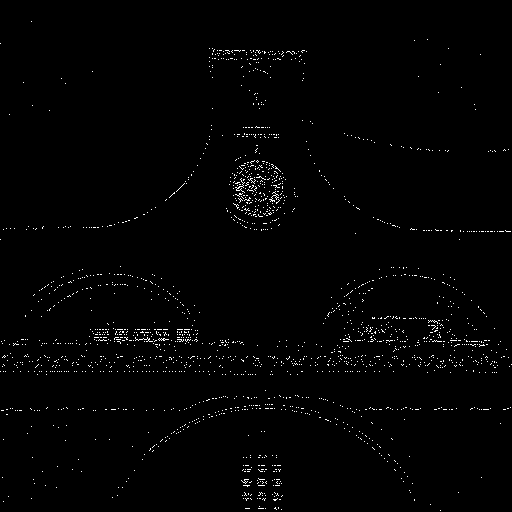} &
    	
    	~ &
    	\includegraphics[width=0.09\textwidth]
    		{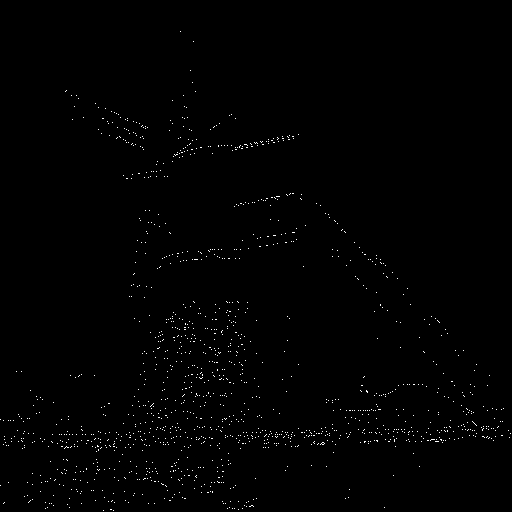} &
    	\includegraphics[width=0.09\textwidth]
    		{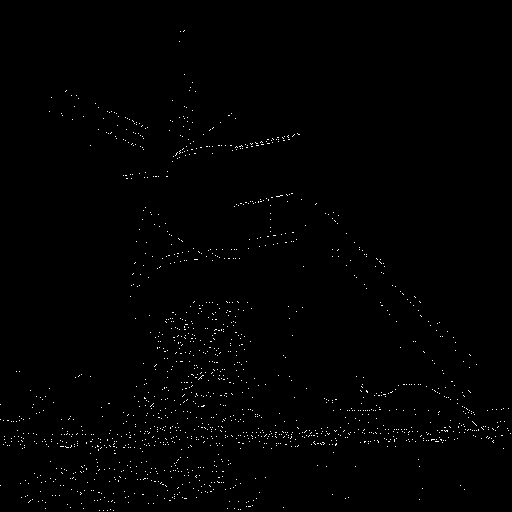} &
    	\includegraphics[width=0.09\textwidth]
    		{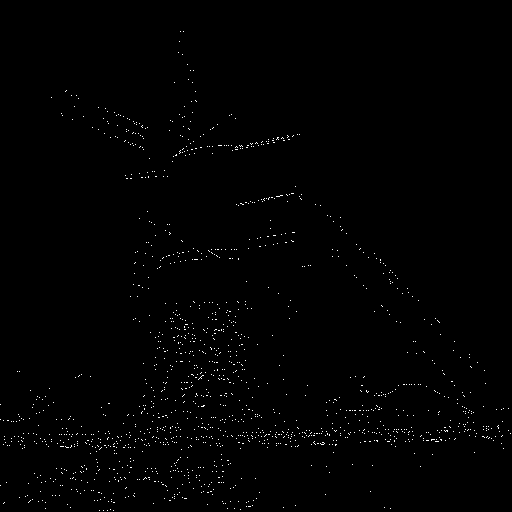} \\
    	
    	\multirow{1}{*}[6ex]{\rotatebox[origin=c]{90}{$2 \times 2$}} &
    	~ &
    	~ &
    	\includegraphics[width=0.09\textwidth]
    		{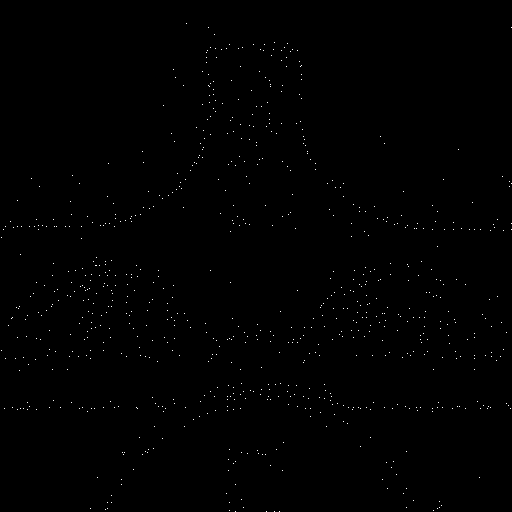} &
    	\includegraphics[width=0.09\textwidth]
    		{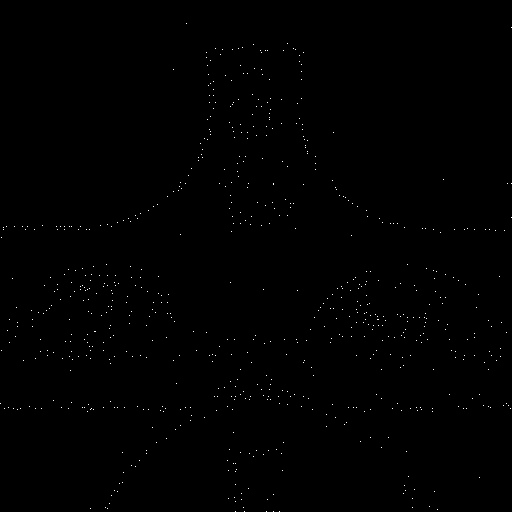} &
    	
    	~ &
    	~ &
    	\includegraphics[width=0.09\textwidth]
    		{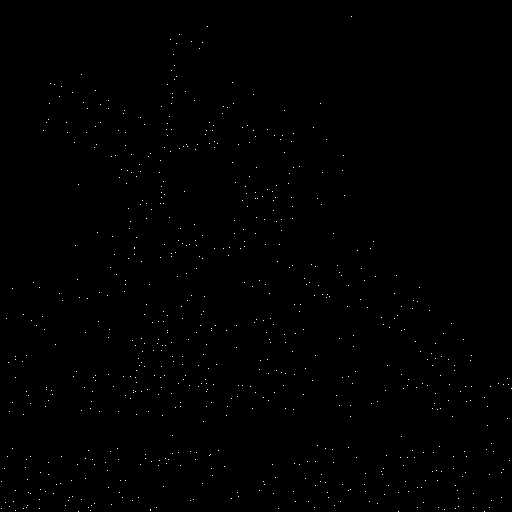} &
    	\includegraphics[width=0.09\textwidth]
    		{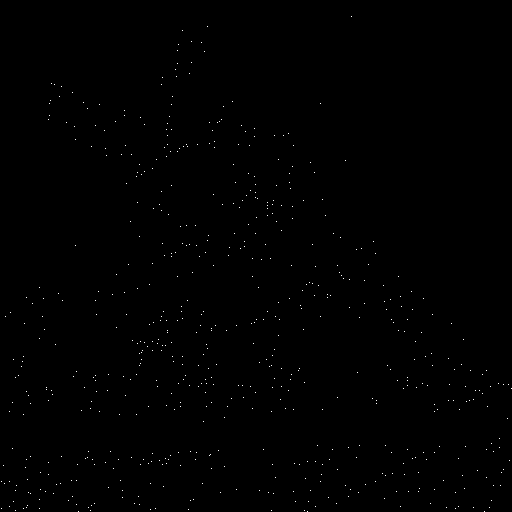} \\
    	
    	\multirow{1}{*}[7ex]{\rotatebox[origin=c]{90}{$16 \times 16$}} &
    	~ &
    	~ &
    	~ &
    	\includegraphics[width=0.09\textwidth]
    		{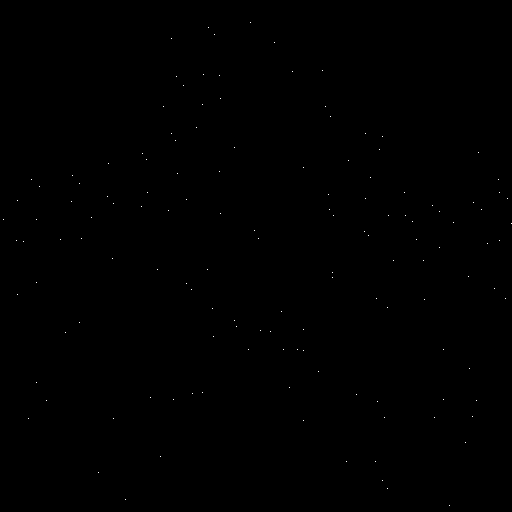} &
    	
    	~ &
    	~ &
    	~ &
    	\includegraphics[width=0.09\textwidth]
    		{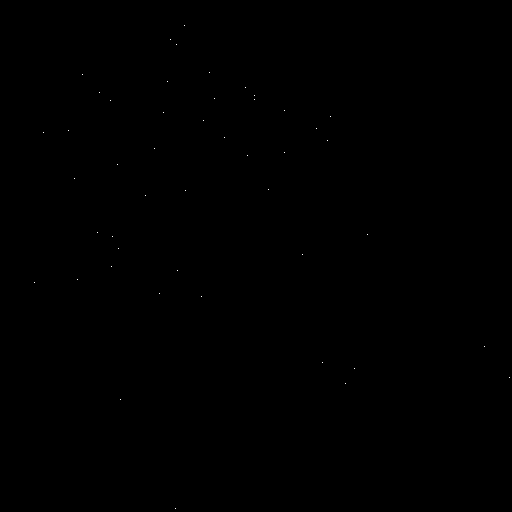} \\
    	
    	~ &
    	MSE:~$32.46$ &
    	MSE:~$27.08$ &
    	MSE:~$24.11$ &
    	MSE:~$\bm{23.25}$ &
    	
    	MSE:~$181.08$ &
        MSE:~$172.11$ &
    	MSE:~$160.10$ &
    	MSE:~$\bm{157.30}$ \\
    	
    \end{tabular}
    }
    \caption{Masks and inpainting results for \emph{elpaso} and 
        \emph{windmill}. The combined mask density is $5\%$. All masks 
        are generated with 30 iterations of our Voronoi densification. 
        Each column gives masks and inpainting result for a different set 
        of feature types. Increasing the variety of feature types improves 
        the reconstruction quality without increasing the total amount
        of information.}
    \label{fig:exp_qualitative}
\end{figure*}

\section{Conclusions and Outlook}
\label{sec:conclusion}

We have shown how an inpainting process can be generalised to incorporate 
an arbitrary set of different feature types in terms of linear constraints. 
While this problem appears to be fundamental, to our knowledge it has not 
been addressed in the inpainting community so far. In view of the simplicity 
and generality of its solution, this is surprising. 
On top of that, we have established the first generic approach for spatial 
and tonal optimisation for multiple different feature types. Finally we have
demonstrated its practical relevance by the quality improvements with a 
novel set of features. The fact that multiscale local averages offer 
superior performance is another interesting observation that may deserve 
further attention in the future. These integrals form a natural 
complement to derivative features. 

In our ongoing work, we are generalising our framework even further
to nonlinear inpainting operators and constraints. Moreover, we also 
envision further improvements w.r.t.~our data optimisation strategies
and a wider range of feature types. Last but not least, we will 
integrate all these concepts in a practical codec that optimises
not only for reconstruction quality but also for coding costs. 

\bibliographystyle{IEEEbib}
\bibliography{myrefs.bib}

\begin{thebibliography}{10}

\bibitem{GL14}
C.~Guillemot and O.~{Le Meur},
\newblock ``Image inpainting: Overview and recent advances,''
\newblock {\em IEEE Signal Processing Magazine}, vol. 31, no. 1, pp. 127--144,
  2014.

\bibitem{GWWB05}
I.~Gali\'c, J.~Weickert, M.~Welk, A.~Bruhn, A.~Belyaev, and H.-P. Seidel,
\newblock ``Towards {PDE}-based image compression,''
\newblock in {\em Variational, Geometric and Level-Set Methods in Computer
  Vision}, N.~Paragios, O.~Faugeras, T.~Chan, and C.~Schn\"orr, Eds., vol. 3752
  of {\em Lecture Notes in Computer Science}, pp. 37--48. Springer, Berlin,
  2005.

\bibitem{SPME14}
C.~Schmaltz, P.~Peter, M.~Mainberger, F.~Ebel, J.~Weickert, and A.~Bruhn,
\newblock ``Understanding, optimising, and extending data compression with
  anisotropic diffusion,''
\newblock {\em International Journal of Computer Vision}, vol. 108, no. 3, pp.
  222--240, July 2014.

\bibitem{MHWT12}
M.~Mainberger, S.~Hoffmann, J.~Weickert, C.~H. Tang, D.~Johannsen, F.~Neumann,
  and B.~Doerr,
\newblock ``Optimising spatial and tonal data for homogeneous diffusion
  inpainting,''
\newblock in {\em Scale Space and Variational Methods in Computer Vision},
  A.~M. Bruckstein, B.~ter Haar~Romeny, A.~M. Bronstein, and M.~M. Bronstein,
  Eds., vol. 6667 of {\em Lecture Notes in Computer Science}, pp. 26--37.
  Springer, Berlin, 2012.

\bibitem{BBG15}
E.-M. Brinkmann, M.~Burger, and I.~Grah,
\newblock ``Regularization with sparse vector fields: From image compression to
  {TV}-type reconstruction,''
\newblock in {\em Scale Space and Variational Methods in Computer Vision},
  {J.-F.} Aujol, M.~Nikolova, and N.~Papadakis, Eds., vol. 9087 of {\em Lecture
  Notes in Computer Science}, pp. 191--202. Springer, Berlin, 2015.

\bibitem{SPHW16}
M.~Schneider, P.~Peter, S.~Hoffmann, J.~Weickert, and E.~Meinhardt-Llopis,
\newblock ``Gradients versus grey values for sparse image reconstruction and
  inpainting-based compression,''
\newblock in {\em Advanced Concepts for Intelligent Vision Systems},
  J.~Blanc-Talon, C.~Distante, W.~Philips, D.~Popescu, and P.~Scheunders, Eds.,
  vol. 10016 of {\em Lecture Notes in Computer Science}, pp. 1--13. Springer,
  Cham, 2016.

\bibitem{Ca88}
S.~Carlsson,
\newblock ``Sketch based coding of grey level images,''
\newblock {\em Signal Processing}, vol. 15, pp. 57--83, 1988.

\bibitem{MBWF11}
M.~Mainberger, A.~Bruhn, J.~Weickert, and S.~Forchhammer,
\newblock ``Edge-based compression of cartoon-like images with homogeneous
  diffusion,''
\newblock {\em Pattern Recognition}, vol. 44, no. 9, pp. 1859--1873, Sept.
  2011.

\bibitem{SCSA04}
A.~Sol\'e, V.~Caselles, G.~Sapiro, and F.~Arandiga,
\newblock ``Morse description and geometric encoding of digital elevation
  maps,''
\newblock {\em IEEE Transactions on Image Processing}, vol. 13, no. 9, pp.
  1245--1262, Sept. 2004.

\bibitem{HMWP13}
S.~Hoffmann, M.~Mainberger, J.~Weickert, and M.~Puhl,
\newblock ``Compression of depth maps with segment-based homogeneous
  diffusion,''
\newblock in {\em Scale Space and Variational Methods in Computer Vision},
  A.~Kuijper, K.~Bredies, T.~Pock, and H.~Bischof, Eds., vol. 7893 of {\em
  Lecture Notes in Computer Science}, pp. 319--330. Springer, Berlin, 2013.

\bibitem{JPW20}
F.~Jost, P.~Peter, and J.~Weickert,
\newblock ``Compressing flow fields with edge-aware homogeneous diffusion
  inpainting,''
\newblock in {\em Proc.~2020 International Conference on Acoustics, Speech, and
  Signal Processing}, Barcelona, Spain, May 2020, pp. 2198--2202.

\bibitem{JPW21}
F.~Jost, P.~Peter, and J.~Weickert,
\newblock ``Compressing piecewise smooth images with the {Mumford--Shah}
  cartoon model,''
\newblock in {\em Proc.~28th European Signal Processing Conference}, Amsterdam,
  Netherlands, Jan. 2021, pp. 511--515.

\bibitem{ZR86}
Y.~Zeevi and D.~Rotem,
\newblock ``Image reconstruction from zero-crossings,''
\newblock {\em IEEE Transactions on Acoustics, Speech, and Signal Processing},
  vol. 34, pp. 1269--1277, 1986.

\bibitem{KLDJ05}
F.~M.~W. Kanters, M.~Lillholm, R.~Duits, B.~J.~P. Jansen, B.~Platel, L.~M.~J.
  Florack, and B.~M. ter Haar~Romeny,
\newblock ``On image reconstruction from multiscale top points,''
\newblock in {\em Scale Space and {PDE} Methods in Computer Vision}, R.~Kimmel,
  N.~Sochen, and J.~Weickert, Eds., vol. 3459 of {\em Lecture Notes in Computer
  Science}, pp. 431--439. Springer, Berlin, 2005.

\bibitem{CCM97}
V.~Caselles, B.~Coll, and {J.-M.} Morel,
\newblock ``Junction detection and filtering,''
\newblock in {\em Foundations of Computational Mathematics}, F.~Cucker and
  M.~Shub, Eds., pp. 23--42. Springer, Berlin, 1997.

\bibitem{WJP11}
P.~Weinzaepfel, H.~J\'egou, and P.~P\'erez,
\newblock ``Reconstructing an image from its local descriptors,''
\newblock in {\em Proc.~2011 IEEE Computer Society Conference on Computer
  Vision and Pattern Recognition}, Colorado Springs, CO, June 2011, pp.
  337--344, IEEE Computer Society Press.

\bibitem{BBBW08}
Z.~Belhachmi, D.~Bucur, B.~Burgeth, and J.~Weickert,
\newblock ``How to choose interpolation data in images,''
\newblock {\em SIAM Journal on Applied Mathematics}, vol. 70, no. 1, pp.
  333--352, 2009.

\bibitem{HSW13}
L.~Hoeltgen, S.~Setzer, and J.~Weickert,
\newblock ``An optimal control approach to find sparse data for {L}aplace
  interpolation,''
\newblock in {\em Energy Minimisation Methods in Computer Vision and Pattern
  Recognition}, A.~Heyden, F.~Kahl, C.~Olsson, M.~Oskarsson, and X.-C. Tai,
  Eds., vol. 8081 of {\em Lecture Notes in Computer Science}, pp. 151--164.
  Springer, Berlin, 2013.

\bibitem{OCBP14}
P.~Ochs, Y.~Chen, T.~Brox, and T.~Pock,
\newblock ``i{P}iano: Inertial proximal algorithm for nonconvex optimization,''
\newblock {\em SIAM Journal on Imaging Sciences}, vol. 7, pp. 1388--1419, 2014.

\bibitem{BLPP17}
S.~Bonettini, I.~Loris, F.~Porta, M.~Prato, and S.~Rebegoldi,
\newblock ``On the convergence of a linesearch based proximal-gradient method
  for nonconvex optimization,''
\newblock {\em Inverse Problems}, vol. 33, no. 5, 2017,
\newblock Article 055005.

\bibitem{APW22}
T.~Alt, P.~Peter, and J.~Weickert,
\newblock ``Learning sparse masks for diffusion-based image inpainting,''
\newblock in {\em Pattern Recognition and Image Analysis}, A.~J. Pinho,
  P.~Georgieva, L.~F. Teixeira, and J.~A. S{\'a}nchez, Eds., Cham, 2022, vol.
  13256 of {\em Lecture Notes in Computer Science}, pp. 528--539, Springer.

\bibitem{KBPW18}
L.~Karos, P.~Bheed, P.~Peter, and J.~Weickert,
\newblock ``Optimising data for exemplar-based inpainting,''
\newblock in {\em Advanced Concepts for Intelligent Vision Systems},
  J.~Blanc-Talon, D.~Helbert, W.~Philips, D.~Popescu, and P.~Scheunders, Eds.,
  vol. 11182 of {\em Lecture Notes in Computer Science}, pp. 547--558.
  Springer, Berlin, Sept. 2018.

\bibitem{DAW21}
V.~Daropoulos, M.~Augustin, and J.~Weickert,
\newblock ``Sparse inpainting with smoothed particle hydrodynamics,''
\newblock {\em SIAM Journal on Imaging Sciences}, vol. 14, no. 4, pp.
  1669--1705, 2021.

\bibitem{Ii62}
T.~Iijima,
\newblock ``Basic theory on normalization of pattern (in case of typical
  one-dimensional pattern),''
\newblock {\em Bulletin of the Electrotechnical Laboratory}, vol. 26, pp.
  368--388, 1962,
\newblock In Japanese.

\bibitem{GMW81}
P.~E. Gill, W.~Murray, and M.~H. Wright,
\newblock {\em Practical Optimization},
\newblock Academic Press, London, 1981.

\bibitem{PS75}
C.~C. Paige and M.~A. Saunders,
\newblock ``Solution of sparse indefinite systems of linear equations,''
\newblock {\em SIAM Journal on Numerical Analysis}, vol. 12, no. 4, pp.
  617--629, 1975.

\bibitem{Sa03}
Y.~Saad,
\newblock {\em Iterative Methods for Sparse Linear Systems},
\newblock {SIAM}, Philadelphia, second edition, 2003.

\end{thebibliography}

\appendix
\renewcommand\thesection{APPENDIX \Alph{section}}
\section{Additional Experiments}

\begin{figure}[h]
    \centering
    \begin{tabular}{cccc}
        \includegraphics[width=0.095\textwidth]
            {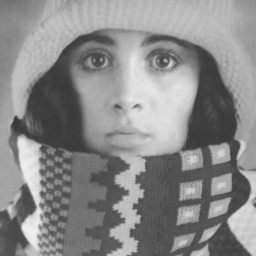} & 
        \includegraphics[width=0.095\textwidth]
            {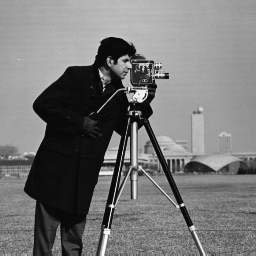} &
        \includegraphics[width=0.095\textwidth]
            {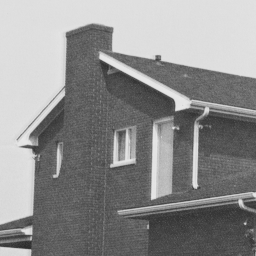} &
        \includegraphics[width=0.095\textwidth]
            {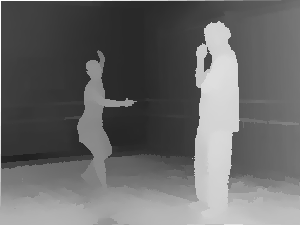} \\
        trui & cameraman & house & ballet
    \end{tabular}
    
    \begin{tabular}{|l|r|r|r|r|}
        \hline
        image & grey values & $\partial_x + \partial_y$ & $2 \times 2$ & $16 \times 16$ \\ 
        \hline
        trui & $38.45$ & $38.51$ & $32.53$ & $31.99$ \\ 
        \hline
        cameraman & $138.27$ & $115.55$ & $110.67$ & $107.90$ \\ 
        \hline
        house & $58.78$ & $53.42$ & $50.69$ & $50.48$ \\ 
        \hline
        ballet & $7.29$ & $2.34$ & $2.29$ & $2.17$ \\ 
        \hline
    \end{tabular}
    \caption{Additional results for greysacale images \emph{trui}, 
        \emph{cameraman}, \emph{house} and depth map \emph{ballet}.
        The table shows MSE values for an increasing number of feature
        types from left to right, similar to \Cref{fig:exp_qualitative}. The
        combined density is again $5\%$ with 30 iterations of our
        spatial optimisation algorithm.}
    \label{fig:additional_results}
\end{figure}

In our experimental evaluation we have used two images to illustrate the performance of our framework (see \Cref{fig:exp_qualitative}). However, those results also generalise to other images, with slight deviations depending on the frequencies of the images. To validate this claim, we present additional experiments in \Cref{fig:additional_results} on standard test images used in the signal processing literature.
The dependence on the frequencies is an effect of the homogeneous diffusion inpainting, rather than the features and the specifics of our framework. 
\ferdinand{Is this claim really true? We have seen that gradient features work better for images with sharp edges.}
The slight increase in MSE for \emph{trui} using greyvalues and derivative features (the $\partial_x + \partial_y$ column) is caused by inaccuracies of the MSE approximation 
used in our spatial optimisation stage (see \Cref{sec:spatial_optimisation}).
Namely, we use the error maps $|\bm{A}_i(\bm{u}-\bm{f})|^2$ as predictors of the 
MSE in order to select mask point locations at a lower computational cost.

\section{Tonal Optimisation}

In \cref{sec:tonal_optimisation} we defined the reconstruction as $\bm{u}=\bm{R}\bm{b}$ in order to formulate the tonal optimisation problem in a concise manner. Here we make the meaning of $\bm{R}\in\mathbb{R}^{Nm\times Nm}$ and $\bm{b}\in\mathbb{R}^{Nm}$ precise. The definition of $\bm{R}$ can be derived starting from (\ref{eq:Lagrange_equation}) as follows:
\begin{align*}
\begin{split}
    \begin{bmatrix} \bm{L} & \bm{A}^T \\ \bm{A} & \bm{0} \end{bmatrix} 
    \begin{bmatrix} \bm{u} \\ \bm{\lambda} \end{bmatrix} 
    &= \begin{bmatrix} \bm{0} \\ \bm{b} \end{bmatrix} 
    \\
    \begin{bmatrix} \bm{u} \\ \bm{\lambda} \end{bmatrix} 
    &=  \begin{bmatrix} \bm{L} & \bm{A}^T \\ \bm{A} & \bm{0} \end{bmatrix}^{+}
    \begin{bmatrix}\bm{0} \\ \bm{I}\end{bmatrix} \bm{b} 
    \\
    \bm{u} =\bm{R}\bm{b} &= \begin{bmatrix} \bm{I} & \bm{0}\end{bmatrix}
    \begin{bmatrix} \bm{L} & \bm{A}^T \\ \bm{A} & \bm{0} \end{bmatrix}^{+}
    \begin{bmatrix}\bm{0} \\ \bm{I}\end{bmatrix}\bm{b} \, .
\end{split}
\end{align*}
In the above $\bm{M}^+$ is the Moore-Penrose pseudoinverse of $\bm{M}$. We do not compute it in practice, instead we solve \Cref{eq:Lagrange_equation} using the SYMMLQ solver and keep only the $\bm{u}$ part. 

The vector $\bm{b}$ can be any vector from $\mathbb{R}^{Nm}$ (in the range of the system matrix) with zeroes at indices corresponding to non-mask pixels. That is, if the mask for the $i$-th feature $\bm{C}_i$ does not 
have a mask pixel at index $j$, i.e.\ $(\bm{C}_i)_{jj}=0$, then the value 
of $\bm{b}$ at index $iN\!+\!j$ is zero: $b_{iN+j} = 0$. The specific choice 
$\bm{b}=\bm{A}\bm{f}$ in (\ref{eq:formulation_compat_constraints}) interpolates $\bm{f}$'s 
grey/colour values, derivatives, and local integrals. However, we can 
theoretically store any values (i.e.\ not necessarily interpolating $\bm{f}$) at mask points. The optimal 
ones in terms of the Euclidean distance minimise $\|\bm{u}-\bm{f}\|=\|\bm{R}\bm{b}-\bm{f}\|$. Tonal optimisation finds those optimal values.

\section{SYMMLQ SOLVER} To solve \Cref{eq:Lagrange_equation} we use the SYMMLQ~\cite{PS75} solver. However, in the literature the algorithm is typically not given explicitly. For completeness we provide pseudocode for the solver in Alg. \ref{alg:symmlq}. Let $\bm{M}\in\mathbb{R}^{n\times n}$ be a symmetric indefinite matrix (e.g.\ the matrix in (\ref{eq:Lagrange_equation})), let $\bm{b}$ be in the range of $\bm{M}$, and we wish to solve $\bm{M}\bm{x} = \bm{b}$, given some initial guess $\bm{x}\in\mathbb{R}^n$ (potentially the zero vector). We denote by $\langle \bm{u},\bm{v}\rangle = \sum_{i=1}^n u_i v_i$ the usual dot product, and by $\|\bm{u}\| = \sqrt{\langle \bm{u}, \bm{u}\rangle}$ the induced norm. Note that in our implementation of (\ref{eq:Lagrange_equation}), all matrix-vector products are performed in a matrix-free fashion in order to maximise efficiency.
\begin{algorithm}
\caption{SYMMLQ Solver} \label{alg:symmlq}
\KwData{$\epsilon > 0, \,\, \bm{M}\in\mathbb{R}^{n\times n}, \,\, \bm{b}\in\mathbb{R}^n, \,\, \bm{x}\in\mathbb{R}^n$}
\KwResult{$\bm{x}\in\mathbb{R}^n \,:\, \bm{M}\bm{x} = \bm{b}$}
\tcc{Initial Lanczos step}
$\bm{u} \gets \bm{M}\bm{x}$;\, $\bm{w} \gets \bm{b} - \bm{u}$;\,
$\beta_0 \gets \|\bm{w}\|$;\, $\bm{w} \gets \bm{w} / \beta_0$\;
$\bm{u} \gets \bm{M}\bm{w}$;\, $\alpha \gets \langle\bm{w},\bm{u}\rangle$;\,
$\bm{u}\gets \bm{u} - \alpha\cdot\bm{w}$\;
\tcc{Initial variables}
$\beta_1\gets \beta_0$;\,$\beta_2 \gets \|u\|$\;
$LQ_{\|\cdot\|} \gets\beta_1$;\,$CG_{\|\cdot\|} \gets\beta_1$;\,
$QR_{\|\cdot\|} \gets\beta_1$;\,$r^L_{1,\|\cdot\|} \gets\beta_1$\;
$\overline{\gamma} \gets \alpha$;\, $\overline{\delta} \gets \beta_2$;\,
$r^L_{2,\|\cdot\|} \gets 0$;\,$S\gets 1$\;
$\overline{T}_{\|\cdot\|^2_F} \gets \alpha^2 + \beta_2^2$;\,$z_{\|\cdot\|^2} \gets 0$;\,
$\overline{\bm{w}}\gets \bm{0}$\;
\While{\textnormal{True}}{
$d \gets \overline{\gamma}$\;
\tcc{Errors and stopping criterion}
\If{$d=0$}{
$d\gets \sqrt{\overline{T}_{\|\cdot\|^2_F}}$\;
}
$LQ_{\|\cdot\|} \gets \sqrt{(r^L_{1,\|\cdot\|})^2 + (r^L_{2,\|\cdot\|})^2}$\;
$QR_{\|\cdot\|} \gets \beta_0 \cdot S$\; $CG_{\|\cdot\|}\gets (\beta_2 \cdot QR_{\|\cdot\|})/|d|$\;
$\overline{\zeta} \gets r^L_{1,\|\cdot\|}/d$\;
\If{$CG_{\|\cdot\|}^2\leq \overline{T}_{\|\cdot\|^2_F} \cdot z_{\|\cdot\|^2} \cdot \epsilon^2 $}{
\textbf{break}\;
}
\tcc{Lanczos step}
$\bm{u} \gets \bm{u}/\beta_2$;\,$\bm{v} \gets \bm{M} \bm{u}$;\,
$\alpha \gets \langle\bm{u},\bm{v}\rangle$\;
$\bm{v} \gets \bm{v} - \alpha\cdot \bm{u} - \beta_2\cdot \bm{w}$;\, $\beta_1 \gets \beta_2$;\,$\beta_2 \gets \|\bm{v}\|$\;
\tcc{Update $\overline{\bm{T}}_{\|\cdot\|^2_F}$ norm}
$\overline{\bm{T}}_{\|\cdot\|^2_F} \gets \overline{\bm{T}}_{\|\cdot\|^2_F}  
            + \alpha^2 + \beta_1^2 + \beta_2^2$\;
\tcc{Next plane rotation}    
$\gamma \gets \sqrt{\overline{\gamma}^2 + \beta_1^2}$;\, 
$cs \gets \overline{\gamma}/\gamma$; \, $sn \gets \beta_1/\gamma$\;
$\delta \gets cs \cdot \overline{\delta} + sn \cdot \alpha$;\,
$\overline{\gamma} \gets sn \cdot \overline{\delta} - cs \cdot \alpha$;\,
$\overline{\delta} \gets -cs \cdot \beta_2$\;
\tcc{Update $\bm{x}$ and $\overline{\bm{w}}$}    
$\zeta \gets r^L_{1,\|\cdot\|}/\gamma$;\, $s\gets cs \cdot \zeta$;\,
$t\gets sn \cdot \zeta$\;
$\bm{x} \gets \bm{x} + s\cdot \overline{\bm{w}} + t\cdot \bm{u}$;\, 
$\overline{\bm{w}} \gets sn\cdot\overline{\bm{w}} - cs \cdot\bm{u}$\;
\tcc{Update variables}
$S\gets sn \cdot S$;\,$z_{\|\cdot\|^2} \gets z_{\|\cdot\|^2} + \zeta^2$\;
$r^L_{1,\|\cdot\|} \gets r^L_{2,\|\cdot\|} - \delta \cdot \zeta$;\,
$r^L_{2,\|\cdot\|} \gets - sn\cdot \beta_2\cdot \zeta$\;
\tcc{Cycle pointers for next iter.}
$\bm{u} \gets \bm{v} \gets \bm{w} \gets \bm{u}$\;
}
\tcc{Transfer to CG point}    
\If{$CG_{\|\cdot\|}\leq LQ_{\|\cdot\|}$}{
$\overline{\zeta} \gets r^L_{1,\|\cdot\|}/d$;\, 
$\bm{x} \gets \bm{x} + \overline{\zeta}\cdot\overline{\bm{w}}$\;
}
\end{algorithm}

%

%
%


\end{document}